\documentstyle[12pt]{article}

\begin{document}

\title{Quantum statistics of ideal gases in confined space}
\author{Wu-Sheng Dai \thanks{Email address: daiwusheng@tju.edu.cn}\\
        {\footnotesize School of Science, Tianjin University, Tianjin
        300072, P. R. China}\\
        and \\
        Mi Xie \thanks{Email address: xiemi@mail.tjnu.edu.cn} \\
        {\footnotesize Department of Physics, Tianjin Normal University,
        Tianjin 300074, P. R. China}
        }
\date{}
\maketitle

\begin{abstract}
In this paper, the effects of boundary and connectivity on ideal gases in
two-dimensional confined space and three-dimensional tubes are discussed in
detail based on the analytical result. The implication of such effects on the
mesoscopic system is also revealed.
\end{abstract}



\section*{1. Introduction}

With the increasing interest in studying the quantum effects in mesoscopic
systems \cite{Jalochowski1,Jalochowski2} which are so small that the boundary
effect gets important and can not be neglected any more, the question of how
to solve the sum of the states precisely, or, how to take the boundary
effect of such system into consideration, is raised naturally. The properties of some 
systems found recently are shape dependent and sensitive to the topology 
\cite{Potempa,Braun,Kravtsov}.

The key problem in statistical mechanics is to solve the sum over all
possible states. In principle, when an ideal gas is confined in finite volume,
the spectrum of single-particle states will be determined by the
configuration of the boundary. In case the boundary is irregular, it is
impossible to get the sum over all possible states exactly. However, if the
mean thermal wavelength of the particles is much shorter than the size of
the system, as an approximation, one can assume that the spectrum of
single-particle states be continuous while the total number of states be
independent of the shape of the boundary and simply proportional to the
volume (in three dimensional case) or the area (in two dimensional case) of
the system. In other words, if the thermal wavelength of particles is very
short in relative to the size of the volume or area which the system
occupies, the effect of boundary configuration on the spectrum can be
ignored. Historically, such an assumption was advocated by radiation theory of
Rayleigh-Jeans. It also aroused great interest to the mathematicians, and
finally was proved by Weyl mathematically \cite{Weyl}.

The above problem in statistical mechanics is related to such an inverse
problem in mathematics, that is, if it is possible to determine the metric
and topological features by the knowledge of the spectrum, which is still an
open question.

After Weyl's leading work\cite{Weyl}, some progress has been made in
mathematics \cite{Kac}. Now we know that, in two dimensions, for the
eigenvalue problem 
\begin{equation}
\frac{1}{2}\nabla ^{2}U+\mu U=0\mbox{  in }{\bf \Omega }\mbox{~~with~}U=0%
\mbox{  on }\Gamma ,  \label{e1}
\end{equation}%
where ${\bf \Omega }$ is the region bounded by curve $\Gamma $, the total
number of eigenstates $N$ can be written as \cite{Kac} 
\begin{equation}
N\sim \sum\limits_{n=1}^{\infty }e^{-\mu _{n}t}\longrightarrow \frac{\Omega 
}{2\pi t}-\frac{L}{4}\frac{1}{\sqrt{2\pi t}}+\frac{1-r}{6}%
,~~~~~(t\rightarrow 0).  \label{e3}
\end{equation}%
Here $\Omega $ is the area of ${\bf \Omega }$, $L$ the length of $\Gamma $,
and $r$ the number of holes in ${\bf \Omega }$, while $\{\mu _{n}\}$ is the
spectrum of eigenvalues of the system. In the short-wavelength limit, Weyl
proved for two-dimensional case that $N\sim \frac{\Omega }{2\pi t}$, i.e.,
the total number of states is proportional to the area of the region, which
is a fundamental assumption in statistical physics.

Nevertheless, Weyl's result omits the last two terms in eq.(\ref{e3})
coming from the perimeter and the number of holes and hence loses part of
the information related to the geometry of the region. Here, the relation
between the perimeter $L$ and the area $\Omega $ contains some information
of the shape of the region while the number of holes $r$ reflects the
connectivity.

In this paper, the effects of the boundary and the connectivity on the
quantum statistics are discussed. In Sec. II, the analytical result with the
geometry effects being considered for ideal Bose and Fermi gases in
two-dimensional space is given, based on which the relevant physical
indications of such effects are revealed. In Sec. III, the quantum statistics
in there-dimensional tubes is discussed. The conclusions are summarized in
Sec. IV while some expressions of useful thermodynamic quantities are
given in Appendix.

\section*{2. Statistics in two-dimensional space}

Obviously, eq.(\ref{e1}) is just the Schr\"{o}dinger equation for free
particles in a two-dimensional container. Based on eq.(\ref{e3}), we have

\begin{equation}
\sum\limits_{s}e^{-\beta \epsilon _{s}}=\frac{\Omega }{\lambda ^{2}}-\frac{1%
}{4}\frac{L}{\lambda }+\frac{1-r}{6},~~~~~(\lambda ^{2}\rightarrow 0),
\label{e5}
\end{equation}%
where $\lambda =h/\sqrt{2\pi mkT}=h\sqrt{\beta }/\sqrt{2\pi m}$ is the mean
thermal wavelength, $\epsilon _{s}$ the energy of a free particle, and the
subscript $s$ labels different states. In macroscopic systems, the area
(volume) of the system is usually large enough so that the influence of
boundary can be ignored. However, it may not be true for some special cases.
The above formula is a more precise approximation which includes the
influence of the shape and the connectivity. It means that the geometry of
the container may result in observable effects in physics. The effects
described by the last two terms of eq.(\ref{e5}) can be expected to be
observable in such cases that, 1) the area (volume) of the system is small,
2) the area (volume) is not small but the boundary of the system is so
complicated that $L\gg \sqrt{\Omega }$ or the system is in multiply
connected space.

For ideal Bose and Fermi gases in two-dimensional space, when the influence
of boundary and connectivity is considered, the grand potential of the
system reads as \cite{Huang}

\begin{equation}
\ln \Xi =\mp \sum\limits_{s}\ln (1\mp ze^{-\beta \epsilon _{s}}).  \label{e7}
\end{equation}%
In this equation and following, the upper sign stands for bosons and the
lower sign for fermions. Substitute eq.(\ref{e5}) into eq.(\ref{e7}) and
expand it as a series of $ze^{-\beta \epsilon _{s}}$, we have

\begin{equation}
\ln \Xi =\sum\limits_{s}\sum\limits_{n}[(\pm 1)^{n+1}\frac{1}{n}(ze^{-\beta
\epsilon _{s}})^{n}].  \label{e8}
\end{equation}%
Of course, such expansion is valid only for $0<z<1$.

By using the eq.(\ref{e5}), we can perform the summation over $s$:%
\begin{equation}
\ln \Xi =\sum\limits_{n}(\pm 1)^{n+1}\frac{z^{n}}{n}(\frac{1}{n}\frac{\Omega 
}{\lambda ^{2}}-\frac{1}{\sqrt{n}}\frac{1}{4}\frac{L}{\lambda }+\frac{1-r}{6}%
).
\end{equation}%
Now, the sum over $n$ gives Bose-Einstein or Fermi-Dirac integral

\begin{equation}
\sum\limits_{n}(\pm 1)^{n+1}\frac{z^{n}}{n^{\sigma }}=\frac{1}{\Gamma
(\sigma )}\int_{0}^{\infty }\frac{x^{\sigma -1}}{z^{-1}e^{x}\mp 1}dx\equiv
h_{\sigma }(z).
\end{equation}%
Here, we introduce a function $h_{\sigma }(z)$ which equals to 
Bose-Einstein integral $g_{\sigma }(z)$ or Fermi-Dirac integral $%
f_{\sigma }(z)$ in Bose or Fermi case, respectively. Finally, the grand
potential can be expressed as

\begin{equation}
\ln \Xi =\frac{\Omega }{\lambda ^{2}}h_{2}(z)-\frac{1}{4}\frac{L}{\lambda }%
h_{\frac{3}{2}}(z)+\frac{1-r}{6}h_{1}(z).  \label{e12}
\end{equation}%
In comparison with the grand potential $\ln \Xi _{free}=\frac{\Omega }{%
\lambda ^{2}}h_{2}(z)$ for ideal gas in two-dimensional free space without
boundary, grand potential eq.(\ref{e12}) depends not only on the area but
also on the boundary as well as the connectivity of the space. The term
which is proportional to the perimeter $L$ of the area, represents the
influence of the boundary while the another term reflects the effect of the
connectivity.

In addition, it is easy to see from eq.(\ref{e12}) that, grand potential of
a system in confined space is less than that in free space since the signs
of these two terms are negative (when $r>1$). It means that the existence of
a boundary and holes tends to reduce the number of states of the system.
This is just because in free space the energy spectrum is continuous while
in confined space the spectrum gets discrete. Thus the number of modes in
confined space is less than that in free space.

The expansion in eq.(\ref{e8}) requires $0<z<1$. In Bose-Einstein
statistics, such a constraint on the fugacity $z$\ is naturally satisfied.
In Fermi-Dirac statistics, we have $0<z<\infty $. Strictly speaking, the
grand potential $\ln \Xi $ can not be expanded as a series of $ze^{-\beta
\epsilon _{s}}$ when $z>1$. However, the first term of $\ln \Xi $ in eq.(\ref%
{e12}) is just the grand potential in free space $\ln \Xi _{free}$ though
the expansion is not rigorous. Since such a treatment provides a correct
result in free space, we may expect that the rest two terms of $\ln \Xi $,
which describe the contributions of boundary and connectivity, are also
valid for $z>1$.

When the thermal wavelength $\lambda $ is much shorter than the size of
container, particle can not feel the shape of boundary. In contrast, for low
frequency waves whose wavelengths are comparable to the size of container,
its spectrum will seriously depend on the boundary. Therefore, when the
ratio between the wavelength and the size of container is extremely small, a
good approximation can be given by the first term of eq.(\ref{e12}) which is
just the result proven by Weyl, and is only related to the area of the
container. However, when the ratio is not negligible, the influence of the
container geometry has to be considered, which is provided by the last two
terms in eq.(\ref{e12}).

One point must be emphasized here. When solving statistics problem, one has
got used to assume that, the number of states be only proportional to the
area and neglect the effect of boundary. Eq.(\ref{e12}) provides a more
complete and precise approximation.

Following general procedures, the relevant thermodynamic quantities of
two-dimensional ideal gas can be achieved easily although the extra two
terms in eq.(\ref{e12}) make the derivation get tedious. Eliminating $z$ in
the two equations

\begin{equation}
\left\{ \matrix{ \displaystyle\frac{P\Omega }{kT}=\frac{\Omega }{\lambda
^{2}}h_{2}(z)-\frac{1}{4}\frac{L}{\lambda
}h_{\frac{3}{2}}(z)+\frac{1-r}{6}h_{1}(z), \cr \displaystyle N=\frac{\Omega
}{\lambda ^{2}}h_{1}(z)-\frac{1}{4}\frac{L}{\lambda
}h_{\frac{1}{2}}(z)+\frac{1-r}{6}h_{0}(z),}\right.  \label{e15}
\end{equation}%
we obtain the equation of state of the ideal gas. The other thermodynamic
quantities are given in Appendix.

The effect of boundary on the thermodynamic quantities(see Appendix) is of
the order of $L/(\sqrt{N}\sqrt{\Omega })$. The factor $L/\sqrt{\Omega }$
reflects some information of the shape of the two-dimensional container. If
the container shape is close to circle or square, the ratio $L/\sqrt{\Omega }
$ is of order $1$. Otherwise, for example, when the shape of the container
is very complex, $L/\sqrt{\Omega }$ can be large and hence the boundary
effect will become significant.

Besides, the boundary effect is suppressed by the factor $1/\sqrt{N}$. In
macroscopic systems, the contribution of boundary is strongly suppressed.
But, the suppression may not be so serious for mesoscopic systems, and so
that is expectable to observe the boundary effect in such systems.

The influence of connectivity is of order $(1-r)/N$, so one may observe the
effect of connectivity when $r$ is comparable to $N$. This result implies
that in some porous media such effects may not be ignored.

\section*{3. Statistics in there-dimensional tubes}

Base on eq.(\ref{e5}), we can also calculate the boundary effect of a
three-dimensional ideal gas in a long tube, of which all transverse cross
sections keep the same.

The $z$-component of the momentum $p_{z}$ is continuous since the length of
the tube $L_{z}$ is made sufficiently large. This allows us to convert the
summation over $p_{z}$ into an integral. Then, we only need to perform the
summations over $p_{x}$ and $p_{y}$, which can be achieved by using the same
procedure above. The grand potential is

\begin{eqnarray}
\ln \Xi &=&\sum\limits_{n}(\pm 1)^{n+1}\frac{z^{n}}{n}\sum\limits_{s}e^{-n%
\beta \epsilon _{s}}  \nonumber \\
&=&\sum\limits_{n}(\pm 1)^{n+1}\frac{z^{n}}{n}\int \frac{dzdp_{z}}{h}%
e^{-n\beta \frac{p_{z}^{2}}{2m}}\sum\limits_{p_{x},p_{y}}e^{-n\beta \frac{%
p_{x}^{2}+p_{y}^{2}}{2m}}.
\end{eqnarray}%
Working out the summations over $p_{x}$ and $p_{y}$ by use of eq.(\ref{e5}),
we obtain

\begin{equation}
\ln \Xi =\frac{L_{z}\Omega }{\lambda ^{3}}h_{\frac{5}{2}}(z)-\frac{1}{4}%
\frac{L_{z}L}{\lambda ^{2}}h_{2}(z)+\frac{1-r}{6}\frac{L_{z}}{\lambda }h_{%
\frac{3}{2}}(z).
\end{equation}%
where $\Omega $ and $L$ denote the area and perimeter of the transverse
cross section of the tube, respectively.

Directly, we can obtain the equation of state%
\begin{equation}
\left\{ \matrix{ \displaystyle\frac{PV}{kT}=\frac{L_{z}\Omega }{\lambda
^{3}}h_{\frac{5}{2}}(z)-\frac{1}{4}\frac{L_{z}L}{\lambda
^{2}}h_{2}(z)+\frac{1-r}{6}\frac{L_{z}}{\lambda }h_{\frac{3}{2}}(z), \cr
\displaystyle N=\frac{L_{z}\Omega }{\lambda
^{3}}h_{\frac{3}{2}}(z)-\frac{1}{4}\frac{L_{z}L}{\lambda
^{2}}h_{1}(z)+\frac{1-r}{6}\frac{L_{z}}{\lambda }h_{\frac{1}{2}}(z).} \right.
\end{equation}%
The thermodynamics quantities are listed in Appendix. The Bose-Einstein
condensation in a three-dimensional tube will be discussed in detail
elsewhere.

A similar analysis indicates that the effects of boundary and connectivity
in three-dimensional tube are of the same order as those in two-dimensional
space.

\section*{4. Conclusions and discussions}

In conclusion, the effects of boundary and connectivity on the statistical
mechanics of ideal gases in two-dimensional confined space and in
three-dimensional tubes are discussed.

In ideal gas theory, one replace the summation over states by an integral
over momentum: $\sum_{s}\longrightarrow V\int \frac{d^{d}p}{h^{d}}$, where $%
d $ denotes the dimension. This replacement is based on the assumption that
the momentum is independent of the boundary of container, which is valid
only when $V\rightarrow \infty $, i.e., there is no boundary. In other
words, such a replacement is equivalent to the assumption that the number of
states is proportional to the area (in two dimensions) or volume (in three
dimensions) of the container. In fact, this treatment is an approximation
that ignores the influence of boundary.

Our analysis shows that, the influence of boundary and connectivity is of
the order $L/(\sqrt{N}\sqrt{\Omega })$ and $(1-r)/N$, respectively. In many
cases, such influences are of order $1/\sqrt{N}$ and $1/N$, therefore,
negligible. However, when these factors $L/\sqrt{\Omega }$ and $r$ are much
bigger than $1$, the effects may be observable. The factor $L/\sqrt{\Omega }%
\gg 1$ corresponds to such cases, for example, the boundary of the region is
very complex or the two-dimensional container is long and narrow. $r\gg 1$
corresponds to the case that there are many holes in the region. In the
meanwhile, the suppression by $1/\sqrt{N}$ or $1/N$\ could also be reduced
in mesoscopic systems, so we can expect that the effect of container
geometry may be observable in mesoscopic scale, especially in containers
with complex boundary or in porous media.

\section*{Acknowledgments}

\bigskip We wish to acknowledge Dr. Yong Liu for his help. This work is
supported partially by Education Council of Tianjin, P. R. China, under
Project No. 01-20102.

\section*{Appendix: Thermodynamic quantities}

\subsection*{1. Thermodynamic quantities in two dimensions}

Internal energy:%
\[
\frac{U}{NkT}=\frac{h_{2}(z)}{h_{1}(z)}\sigma _{2}-\frac{1}{8\sqrt{N}}\frac{L%
}{\sqrt{\Omega }}\frac{h_{3/2}(z)}{h_{1}^{1/2}(z)}\sigma _{2}^{1/2}. 
\]%
Free energy:

\[
\frac{F}{NkT}=\ln z-\left( \frac{h_{2}(z)}{h_{1}(z)}\sigma _{2}-\frac{1}{4%
\sqrt{N}}\frac{L}{\sqrt{\Omega }}\frac{h_{3/2}(z)}{h_{1}^{1/2}(z)}\sigma
_{2}^{1/2}+\frac{1-r}{6N}h_{1}(z)\right) . 
\]%
Entropy:

\[
\frac{S}{Nk}=2\frac{h_{2}(z)}{h_{1}(z)}\sigma _{2}-\ln z-\frac{3}{8\sqrt{N}}%
\frac{L}{\sqrt{\Omega }}\frac{h_{3/2}(z)}{h_{1}^{1/2}(z)}\sigma _{2}^{1/2}+%
\frac{1-r}{6N}h_{1}(z). 
\]%
Specific heat:

\begin{eqnarray*}
\frac{C_{V}}{Nk} &=&\sigma _{2}\,\left( 2\frac{h_{2}(z)}{h_{1}(z)}-\eta _{2}%
\frac{h_{1}(z)}{h_{0}(z)}\right) \\
&&-\frac{1}{\sqrt{N}}\frac{L}{\sqrt{\Omega }}\sigma _{2}^{1/2}\left( \frac{3%
}{16}\frac{h_{3/2}(z)}{h_{1}^{1/2}(z)}-\frac{1}{8}\eta _{2}\frac{%
h_{1}^{1/2}(z)h_{1/2}(z)}{h_{0}(z)}\right) ,
\end{eqnarray*}%
where%
\[
\sigma _{2}=\left( \frac{1-\frac{1-r}{6N}h_{0}(z)}{\sqrt{1+\frac{1}{64N}%
\frac{L^{2}}{\Omega }\,\frac{h_{1/2}^{2}(z)}{h_{1}(z)}-\frac{1-r}{6N}h_{0}(z)%
}-\frac{1}{8\sqrt{N}}\frac{L}{\sqrt{\Omega }}\frac{h_{1/2}(z)}{h_{1}^{1/2}(z)%
}}\right) ^{2}, 
\]%
\[
\eta _{2}=\frac{1-\frac{1}{8\sqrt{N}}\frac{L}{\sqrt{\Omega }}\frac{h_{1/2}(z)%
}{h_{1}^{1/2}(z)}\frac{1}{\sigma _{2}^{1/2}}}{1-\frac{1}{4\sqrt{N}}\frac{L}{%
\sqrt{\Omega }}\frac{h_{1}^{1/2}(z)h_{-1/2}(z)}{h_{0}(z)}\frac{1}{\sigma
_{2}^{1/2}}+\frac{\,1-r}{6N}\frac{h_{1}(z)h_{-1}(z)}{h_{0}(z)}\frac{1}{%
\sigma _{2}}}. 
\]%
The following relation is used to calculate $C_{V}$%
\[
\frac{\partial z}{\partial T}=-\frac{z}{T}\frac{h_{1}(z)}{h_{0}(z)}\eta
_{2}. 
\]

\subsection*{2. Thermodynamic quantities in three dimensions}

Internal energy:

\[
\frac{U}{NkT}=\frac{3}{2}\frac{h_{5/2}(z)}{h_{3/2}(z)}\sigma _{3}-\frac{1}{%
4N^{1/3}}\frac{L_{z}^{1/3}L}{\Omega ^{2/3}}\frac{h_{2}(z)}{h_{3/2}^{2/3}(z)}%
\sigma _{3}^{2/3}+\frac{1-r}{12N^{2/3}}\frac{L_{z}^{2/3}}{\Omega ^{1/3}}%
h_{3/2}^{2/3}(z)\sigma _{3}^{1/3}. 
\]%
Free energy: 
\begin{eqnarray*}
\frac{F}{NkT} &=&\ln z-[\frac{h_{5/2}(z)}{h_{3/2}(z)}\sigma _{3}-\frac{1}{4}%
\frac{1}{N^{1/3}}\frac{L_{z}^{1/3}L}{\Omega ^{2/3}}\frac{h_{2}(z)}{%
h_{3/2}^{2/3}(z)}\sigma _{3}^{2/3} \\
&&+\frac{1}{N^{2/3}}\frac{1-r}{6}\frac{L_{z}^{2/3}}{\Omega ^{1/3}}%
h_{3/2}^{2/3}(z)\sigma _{3}^{1/3}].
\end{eqnarray*}%
Entropy:

\[
\frac{S}{Nk}=\frac{5}{2}\frac{h_{5/2}(z)}{h_{3/2}(z)}\sigma _{3}-\ln z-\frac{%
1}{2N^{1/3}}\frac{L_{z}^{1/3}L}{\Omega ^{2/3}}\frac{h_{2}(z)}{%
h_{3/2}^{2/3}(z)}\sigma _{3}^{2/3}+\frac{1-r}{4N^{2/3}}\frac{L_{z}^{2/3}}{%
\Omega ^{1/3}}h_{3/2}^{2/3}(z)\sigma _{3}^{1/3}. 
\]%
Specific heat:

\begin{eqnarray*}
C_{V} &=&Nk\{\sigma _{3}[\frac{15}{4}\frac{h_{5/2}(z)}{h_{3/2}(z)}-\frac{9}{4%
}\eta _{3}\frac{h_{3/2}(z)}{h_{1/2}(z)}] \\
&&-\frac{1}{N^{1/3}}\frac{LL_{z}^{1/3}}{\Omega ^{2/3}}\sigma _{3}^{2/3}[%
\frac{h_{2}(z)}{2h_{3/2}^{2/3}(z)}-\frac{3}{8}\eta _{3}\frac{%
h_{3/2}^{1/3}(z)h_{1}(z)}{h_{1/2}(z)}] \\
&&+\frac{1-r}{6}\frac{1}{N^{2/3}}\frac{L_{z}{}^{2/3}}{\Omega ^{1/3}}\sigma
_{3}^{1/3}[\frac{3}{4}(1-\eta _{3})h_{3/2}^{2/3}(z)]\},
\end{eqnarray*}%
where%
\[
\sigma _{3}=\left( 1-\frac{L_{z}L}{\Omega }\frac{1-r}{72N}\frac{%
h_{1}(z)h_{1/2}(z)}{h_{3/2}(z)}+\frac{L_{z}^{2}}{\Omega }\frac{\left(
1-r\right) ^{3}}{2916N^{2}}\frac{h_{1/2}^{3}(z)}{h_{5/2}(z)}\right) ^{-1}\xi
_{1}^{-3}, 
\]%
\[
\eta _{3}=\frac{1-\frac{1}{6N^{1/3}}\frac{L_{z}^{1/3}L}{\Omega ^{2/3}}\frac{%
h_{1}(z)}{h_{3/2}^{2/3}(z)}\frac{1}{\sigma _{3}^{1/3}}+\frac{1-r}{18N^{2/3}}%
\frac{L_{z}^{2/3}}{\Omega ^{1/3}}\frac{h_{1/2}(z)}{h_{3/2}^{1/3}(z)}\frac{1}{%
\sigma _{3}^{2/3}}}{1-\frac{1}{4N^{1/3}}\frac{L_{z}^{1/3}L}{\Omega ^{2/3}}%
\frac{h_{3/2}^{1/3}(z)h_{0}(z)}{h_{1/2}(z)}\frac{1}{\sigma _{3}^{1/3}}+\frac{%
1-r}{6N^{2/3}}\frac{L_{z}^{2/3}}{\Omega ^{1/3}}\frac{%
h_{3/2}^{2/3}(z)h_{-1/2}(z)}{h_{1/2}(z)}\frac{1}{\sigma _{3}^{2/3}}}, 
\]%
\[
\xi _{1}=\xi _{2}-\frac{1}{\xi _{2}}\frac{1}{12N^{1/3}}\frac{L_{z}^{1/3}L}{%
\Omega ^{2/3}}\frac{h_{1}(z)}{h_{3/2}^{2/3}(z)}\xi _{3}^{1/3}+\frac{1-r}{%
18N^{2/3}}\frac{L_{z}^{2/3}}{\Omega ^{1/3}}\frac{h_{1/2}(z)}{h_{3/2}^{1/3}(z)%
}\frac{1}{\xi _{4}^{1/3}}, 
\]%
\[
\xi _{2}=\left( \frac{1}{2}+\frac{1}{2}\sqrt{1+\frac{1}{432N}\frac{L_{z}L^{3}%
}{\Omega ^{2}}\frac{h_{1}^{3}(z)}{h_{3/2}^{2}(z)}\xi _{3}}\right)
^{1/3},~~\xi _{3}=\frac{\xi _{5}^{3}}{\xi _{4}^{2}}, 
\]%
\begin{eqnarray*}
\xi _{4} &=&1-\frac{1-r}{72N}\frac{L_{z}L}{\Omega }\frac{h_{1}(z)h_{1/2}(z)}{%
h_{3/2}(z)}+\frac{\left( 1-r\right) ^{3}}{2916N^{2}}\frac{L_{z}^{2}}{\Omega }%
\frac{h_{1/2}^{3}(z)}{h_{3/2}(z)}, \\
\xi _{5} &=&1-\frac{\left( 1-r\right) ^{2}}{27N}\frac{L_{z}}{L}\frac{%
h_{1/2}^{2}(z)}{h_{1}(z)}.
\end{eqnarray*}%
We also have

\[
\frac{\partial z}{\partial T}=-\frac{3}{2}\frac{z}{T}\frac{h_{3/2}(z)}{%
h_{1/2}(z)}\eta _{3}. 
\]

\end{document}